# LONG RESPONSE TO SCHEUER-YARIV: "A CLASSICAL KEY-DISTRIBUTION SYSTEM BASED ON JOHNSON (LIKE) NOISE – HOW SECURE?", PHYSICS/0601022

L. B. KISH

*Department of Electrical and Computer Engineering, Texas A&M University, College Station, TX 77843-3128, USA*

(Original long version, 1 February 2006)
(Updated with brief version, July 27, 2006)

This is the longer (partially unpublished) version of response; the shorter version (http://arxiv.org/abs/physics/0605013) is published in Physics Letters A. We point out that the claims in the comment-paper of Scheuer and Yariv are either irrelevant or incorrect. We first clarify what the security of a physically secure layer means. The *idealized* Kirchoff-loop-Johnson-like-noise (KLJN) scheme is totally secure therefore it is more secure than idealized quantum communication schemes which can never be totally secure because of the inherent noise processes in those communication schemes and the statistical nature of eavesdropper detection based on error statistics. On the other hand, with sufficient resources, a *practical/non-ideal* realization of the KLJN cipher can arbitrarily approach the idealized limit and outperform even the idealized quantum communicator schemes because the non-ideality-effects are determined and controlled by the design. The cable resistance issue analyzed by Scheuer and Yariv is a good example for that because the eavesdropper has insufficient time window to build a sufficient statistics and the actual information leak can be designed. We show that Scheuer's and Yariv's numerical result of 1% voltage drop supports higher security than that of quantum communicators. Moreover, choosing thicker or shorter wires can arbitrarily reduce this voltage drop further; the same conclusion holds even according to the equations of Scheuer and Yariv.

**First, we present the he brief version of the response which in press at Phys. Lett. A (July 27, 2006). After the brief version a longer, older and more detailed version is shown. Note, the brief version contains simulation results, too.**

**Brief version.** The comment-paper by Scheuer and Yariv (Sch-Y) [1,2] attempts to show that the Kirchhoff-loop-Johnson-noise (*KLJN*) [2-8] classical communicator has limited security. Clarification of the issues of real security is indeed important especially because the KLJN system has recently became *network-ready* [6]. This new property [6] opens a large scale of practical applications because the KLJN cipher can be installed as a computer card [6], similarly to Eternet cards.

In this response, we focus on the most significant issues but additional arguments have been published in [9]. Sch-Y [1,2] claim that the first *KLJN* paper [3] has basic flaws and that the *KLJN* cipher is not secure. However, their arguments are incorrect and/or irrelevant [9], or already published [8]. Because these kind of mistaken claims about the security of physical secure layers arise due to mixing requirements of idealized cipher schemes with practical design issues, first we clarify what do idealized and practical security mean. We have recently published similar considerations in the paper about the natural immunity of the *KLJN* cipher against the *man-in-the-middle* attack [5]. Work is in hand about the mathematical analysis of the practical security and design aspects of the *KLJN* cipher [7]. In our response below we shall use some preliminary results of [7].

First of all, let us consider what *idealized (theoretical) total security* and *practical total security* of physical secure layers mean.
*i) Total security of the idealized model* [4]. This holds when the mathematical model of the idealized physical system shows an unconditional security. It means that the eavesdropper can extract zero information within the mathematical framework of the model; or if she is able to extract information, she disturbs the channel and she will be discovered. The KLJN cipher satisfies these conditions: in the exact mathematical model, the passive eavesdropper extracts zero information and the invasive eavesdropper extracts one bit of information until the eavesdropping is discovered [3,4].

*ii) Total security of the practical situation* [4]. Since a real physical system is always more complex than idealized mathematical models, no practical system can be totally secure. For example, regarding quantum communication, ideal single-photon-source, noise-free channel and noise-free detectors do not exist and any of these non-idealities compromise total security. Still, we can talk about *total practical security* if the security can arbitrarily be increased by unlimited investment in the enhancement of the system. For example, the wire resistance in the KLJN system is also a non-ideality factor [8] however it can be arbitrarily reduced by using ticker wires or shorter connections, see below.

*iii) Information leak at idealized and practical situations*. In idealized quantum communication systems, information of the order of a few % of the number of transmitted bits can be extracted, for example if the eavesdropper randomly extracts a small fraction of photons [11,12], clone them (with ~70% fidelity), sends one photon back to the channel, and extracts the information from the remaining photon. This process will cause only a negligible change of the error rate in the channel so the quantum eavesdropper detection methods will not be able to detect the eavesdropping [10-12]. For that reason practical quantum communicators [12] must use *privacy amplification* technique [10-12], which is a software-based tool, extracting a short key with *guaranteed negligible information leak* from a long key with *possibly strong information leak*. Because the privacy amplification can be used in any secure communication system with any raw key, the fair comparison of the information leak of physical-secure-layer-type communicators requires the comparison of the information leaks of the raw bits. We will show that the KLJN cipher's information leak can easily be much less than that of idealized quantum communicators and this claim is supported even by Sch-Y's results when we analyze their implications. Now, let us deal with the main claims of Sch-Y in their comment paper.

**1.** In the first section Sch-Y state that our analysis in [3] *"contains a basic flaw"* because *"it completely ignores the finite propagation time"* between the sender and receiver. This statement is incorrect. The analysis in [3] is carried out in the so-called *quasi-static limit* of Maxwellian electrodynamics [13] guaranteeing that the *voltage and current along a wire are constant* and that excludes any propagation delay effects. This is expressed by Eq. (9) in [3] and the related text:

$$f_{\max} L \ll c \tag{1}$$

where $f_{\max}$ is the highest frequency component of the voltage and current, $L$ is the cable length (range of communication) and $c$ is the propagation velocity. *This equation implies not only a necessary condition for the security but also the necessary condition that our Kirchoff-loop circuit model, the idealized mathematical model of the cipher, holds* because this condition is the base of the theory of electronic circuits with discrete components. With this criticism, Sch-Y have missed to realize that the high-frequency limit posed by Eq. (1) in the *frequency domain* is equivalent to the quasi-static limit in the *time domain*. Transformation of limits between the time and frequency domain are basic and well-known tools due to Fourier-theory so the reader was supposed to have and imply this knowledge.

**2.** In Section 2, Sch-Y analyze the limit of high-frequency bandwidth (short correlation time) of noise. This limit was excluded in [3] by the Eq. (1) shown above because then the KLJN cipher naturally fails to function; therefore Sch-Y's analysis is irrelevant for [3]. Fortunately, Sch-Y also arrive at the same conclusion that the KLJN cipher is useless in this limit, thus there is no contradiction between Sch-Y [1,2] and [3] in this respect.

**3.** At the beginning of Section 3 of [1,2], Sch-Y claim rightly that if the ciphers use fast switches then propagation time effects will still occur, even with low-frequency noise, and the cipher will be non-secure. Though it is correct, this claim is *again irrelevant* because due to Fourier-theory, it violates directly Eq. (1) shown above, so this situation was excluded in [3]. As it is well-known and mentioned above, short time scale and fast time processes in the time domain correspond to high frequency scale and high frequency components in the frequency domain consequently fast switching and their transients produce high-frequency components in the channel. Therefore, a practical realization of the KLJN cipher will naturally need line filters, which are low-pass filters at the two ends of the line [4], see Figure 1. This is also necessary to defend against possible attacks by high-frequency probing signals. Of course, these are all practical realization problems [4,7] and as such they were out of the scope of [3].

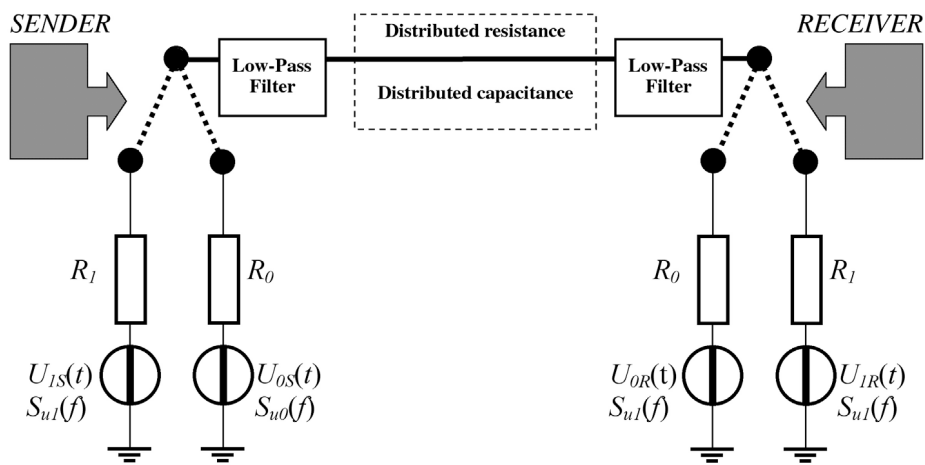

**Figure 1.** Toward the modeling and realization of practical KLJN ciphers.

**4.** Finally, in the rest of Section 3 of [1,2], Sch-Y analyze the practical problem of non-zero wire resistance. Note, this security problem was first pointed out by Janos Bergou [8]. At certain wire length and specific resistance, their numerical results for the relative difference of the mean-square (MS) voltages at the two ends of the line, in the case of secure bit exchange, is 1%. Then Sch-Y claim that this 1% drop of the MS voltage *"can easily be detected and allow Eve to determine Alice and Bob's selection of resistors"*. Though we accept the 1% drop of the MS voltage as a realistic practical goal [7] we disagree with Sch-Y's claim that the eavesdropper can easily detect this 1% drop.

With the very same voltage drop, we have carried out a model study [7] of the distribution functions of the voltages, currents and the drop of the MS voltage for $R_1 / R_0 = 10$, with a *linear full-wave detector* [14] and clock period

$$\tau_c = 3 / f_{\max} .  \qquad (2)$$

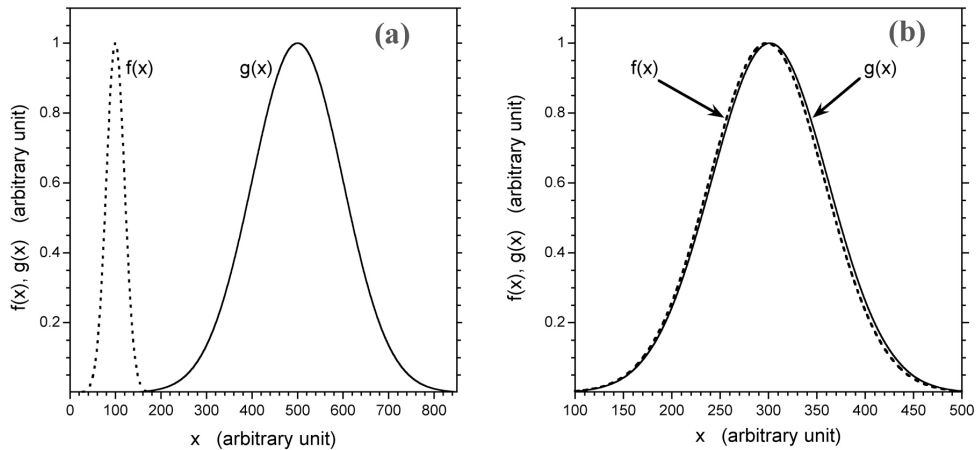

**Figure 2.** Model study of distribution functions [7]. (a): Amplitude distribution functions sampled by the sender and receiver. (b): Amplitude distribution functions sampled by the eavesdropper at the two ends of the wire.

Eq. (2) results in *relative standard deviation* 0.2 of the voltage and current statistics [14]. The results are summarized in Figure 2. Note, only the relative positions and the shape of the curves have meaning, not the actual x and y values. During the clock period, due to Eq. (2), the time is enough only for a few statistically independent sampling of these distribution functions. This sampling is enough for the sender and the receiver, see Figure 2 (a), to decide between the two functions with 0.3% error rate [7]. However the eavesdropper, who measures the voltage drop, has to decide between the two situations by sampling the f(x) and g(x) density functions given in Fig. 2 (b) and that must be done with the same small number of independent samples. The *characteristics width* (standard deviation) of these curves (20% of the peak's x coordinate) is 20 times greater than the difference of the locations of the x coordinates of the peaks (1%). The eavesdropper's task seems to be hopeless by the naked eye however, by using proper statistical tools, she can still extract some information. A deeper analysis based on *Shannon's channel coding theorem* [7] concludes that in this case the *upper limit of information leak* is 0.7% of the transmitted bits. This is close to but less than the information leak of quantum communicators without privacy amplifier software (see above). Thus Sch-Y's 1% drop of the MS voltage yields a lower information leak than that of quantum communicators.

To illustrate how the security can arbitrarily be increased, let us suppose that we increase the wire diameter by a factor of 10. Then, according to Sch-Y's Eq. (12), the relative voltage drop will decrease by a factor of 100 which makes the difference of the location of the x coordinates of the peaks 100 times less than it is presently in Figure 2 (b). This is an impressive improvement making it virtually impossible for the eavesdropper to extract useful information. These considerations illustrate why the KLJN cipher can be designed so that its practical realization is much closer to the limit of *unconditional security* than even the idealized quantum communicators.

**Longer and older (unpublished) version with additional information:**

**1. Introduction**

Scheuer and Yariv have published a preprint [1] addressing some of the practical aspects of the Kirchhoff-loop-Johnson-noise (*KLJN*) cipher of Kish [2-4] and claims that the *KLJN* paper has basic flaws and that the *KLJN* cipher is not secure. Unfortunately, the arguments are incorrect and/or irrelevant [3], or already published [5]. Because these kind of mistaken claims about the security of physical secure layers arise due to mixing requirements of idealized cipher schemes with practical design issues, first we clarify what an idealized and a practical security means. These thoughts have already been published recently [4] in a paper about the natural immunity of the *KLJN* cipher against the *man-in-the-middle* attack. Another paper with general security and design aspects of the *KLJN* cipher will soon be completed [6].

**2. General clarifications about the issue of security of physical secure layers**

In secure communication, any one of the following cases implies absolute security, however the relevant cases for physical secure layers are points 3 and 4 below:

1. The eavesdropper cannot physically access the information channel.

2. The sender and the receiver have a shared secret key for the communication.

3. *The eavesdropper has access and can execute measurements on the channel but the laws of physics do not allow extracting the communicated information from the measurement data.*

4. *The eavesdropper can extract the communicated information however, when that happens, it disturbs the channel so that the sender and receiver discover the eavesdropping activity.*

Keeping points 3 and 4 in mind, we can classify the research of physical secure layers as follows:

*i) Absolute security of the idealized situation* [4]. This is the most fundamental scientific part of the research and the mathematical model of the idealized physical system is developed and tested. The basic question is that how much information can be extracted from the data by the physical measurements allowed in the idealized situation? Paper [2] analyses the KLJN cipher in this respect. The conclusion is that the KLJN cipher provides a greater security level than quantum communicators due to the *robustness of classical information* and the properties of *Fluctuation-Dissipation* mechanism of statistical physics.

*ii) Absolute security of the practical situation* [4]. Though a lot of energy has been devoted to such questions in private discussions and seminar, only generic comments have been published [4,5] and a thorough analysis of the practical security design aspects of the *KLJN* cipher is still missing or it is under preparation [6]. This part of the research requires an interdisciplinary effort including the fields of physics, engineering and data security. Because no real system can totally match the physical properties of the ideal mathematical model system, this kind of absolute security *does not exist in reality*; it is rather approached by *only practically absolute* security. For example, in quantum communication, we have no ideal single photon source, no noise-free channel, and no noise-free detectors, and any of these deficiencies compromise absolute security. Similarly, with the *KLJN* cipher, cable resistance [5] and cable capacitance can cause information leak because the eavesdropper can execute measurements along the cable. This effect can be controlled and minimized by the particular design and choosing proper driving resistances and noise bandwidth that prohibit to make an acceptable statistics about the deviation of the noise strength along the cable within the clock period. Similarly, fast switching of the resistors can violate the *no-wave bandwidth rule* described in [1] (Eq. 9) but this problem can be avoided by using slow switches and/or filters at the line input.

iii) If the code is broken, how many bits can be extracted by the eavesdropper before she is discovered due to the disturbance of the channel? This question can also be treated at both the idealized-fundamental level and at the practical one. Some answers for the idealized case: RSA: infinite number of bits; Idealized Quantum communicator: 20 - 10000 bits; Idealized *KLJN* cipher: 1 bit.

## 3. Response/denial to/of the comments in the Scheuer and Yariv preprint [1]

The analysis in [1] addresses the practical aspect of the Kirchhoff-loop-Johnson-noise (*KLJN*) cipher of Kish. Unfortunately, the arguments are incorrect and/or irrelevant, or already published.

*a)* The study criticizes the security of the *KLJN* cipher by assuming a situation with wide bandwidth, that is short correlation time of the noise (Section 2); fast switches (Section 3); and resistive lines (Section 3).

*b)* First of all, their analysis is focused on practical applications thus it is irrelevant for the *idealized* (mathematical) case of the *KLJN* cipher; see in the *KLJN* papers [2,4]. At the idealized case, neither of these effects is present and the practical case can arbitrarily approach the idealized situation depending on the available investment for components, such as cables, etc. A physically secure layer's security analysis should always start with the idealized case where the fundamental question is the total security of the mathematical model. The original *KLJN* paper [2] considers only one practical issue, the wave propagation, in order to estimate the ultimate bandwidth limit of the idealistic model. The second *KLJN* paper discusses the natural immunity of *KLJN* cipher against the man-in-the-middle attack. However, by no means papers [2,4] contain any practical design issue. Therefore the manuscript [1] and paper [2,4] are like apple and orange. *In conclusion, the Scheuer-Yariv manuscript [1] does not identify any security hole in the idealized (mathematical) case of the KLJN cipher.*

*c)* Section 1 mentions the case of *short correlation time* as a *basic flaw* in paper [2] and Section 2 is dedicated to the same issue: the short correlation time issue. However, the short correlation time limit is explicitly forbidden even by the first *KLJN* paper [2]; see for example the beginning part of Section 5 and Eq. 9. A relevant citation from that part of paper [2] is: "There are two factors limiting speed and range. Firstly, Kirchoff's laws should hold." And then the *no-wave condition* Eq. 9 provides the upper limit of voltage bandwidth in the wire as:

$$f_{\max} L << c \ . \qquad (1)$$

*As it is well known due to Fourier analysis basics, Eq (9) in the frequency domain is equivalent to a lower limit of correlation time in the time domain.* Because of the lower limit of the noise correlation time is the reciprocal of $f_{\max}$, this equation directly excludes the situation with short correlation times, and all the wave reflection argumentations of [1] in Sec 2. Therefore, the "basic flaw" comment in Section 1 and the whole Section 2 are completely invalid/irrelevant.

*d)* In the first part of Sec 3, the manuscript [1] discusses the case of long correlation times, and that is in principle fine. However if the authors suppose there that the *switches are fast* switches, which would obviously violate Eq. 9 because of the high-frequency products violating the *no-wave condition* given by Eq. 9 as well as the above-cited requirement in paper [2] and naturally would cause transient generation, propagation and reflection. This is a valid problem, thus Eq. 9 should be interpreted for any procedure affecting the voltage in the wire. This and similar questions have intensively been discussed among scientists when the *KLJN* cipher hit the media in December 2005 ; see the names in the Acknowledgement. However, these kinds of problems are practical design/engineering issues and these phenomena can be controlled by choosing the details of the particular design. For example, the wire resistance, which is an important issue of non-ideality, can be controlled by the thickness of the wire and that is limited only by the price. Thus the existence of these engineering problems does not challenge the security of the *idealized* cipher. In a realized *KLJN* cipher [2,6], *slow switches* and *low-pass filters* would guarantee the same bandwidth limit for the switching transients as that for the noise bandwidth (reciprocal of the correlation time). Therefore, the security of the realized system can *arbitrarily approach* that of the idealized system, depending on the level of invested efforts and available funds.

*e)* In the second half of Section 3 of preprint [1], it is pointed out that the non-zero resistance of the cable also represents a security leak. This argument is again a practical

*L.B. Kish*

one and it is irrelevant for the idealized case [2,4]. Moreover, this argument about practical realizations is *already known and published*; it was first pointed out by Janos Bergou [5] in the Science Magazine feature [3]. Again, in a practical system, precautions will be made to use appropriately small wire resistance to prohibit obtaining usable statistics about the drop of the noise strength along the cable [2].

f) Finally, at the end of Section 3 of [1], the authors claim that classical key-distribution systems cannot be un-conditionally secure, dislike quantum key-distribution systems. This is an incorrect statement. We have shown in [2,4] and mentioned above that, at idealized conditions, the eavesdropper is discovered after extracting a *single bit* from the *KLJN* cipher. However, the number of extractable bits in an *idealized* quantum communicator is much larger (typically a few thousand bits) because a reasonable statistics of the error rate has to be achieved to detect the eavesdropper. Moreover, as we mentioned above, the KLJN cipher is naturally immune against the *man-in-the-middle attack* [4] which type of attack is a serious concern in most of quantum communicators. On the other hand, any *non-ideal* physical layer can only approach the total security and this statement holds for both the quantum and the *KLJN* ciphers, respectively. The quantum eavesdropper can hide in the statistical noise and that is the reason why marketed quantum communicators need a "privacy amplifier", which is a software tool, to decrease the information leak during communication. A very cheap realization (with thin wires) of the *KLJN* cipher may need a similar privacy amplifier to keep wiring costs at a very cheap level. However, because of the very low costs of the installation of the *KLJN* cipher, there is a good chance that such a tool can be avoided by the proper design of the cipher, see below. Thus, due to the robustness of classical information and the properties of *Fluctuation-Dissipation* mechanism of statistical physics, the *KLJN* cipher is superior to know quantum communicator solutions, including the level of security provided.

In conclusion, from the three arguments of manuscript [1], one is incorrect/irrelevant (Section 2) because violates Eq (9); another one (fast switches, beginning of Section 3) is correct but irrelevant for the idealized case moreover any practical design will automatically take care of it by expanding Eq (9) for the switches; and the last one (cable resistance, second half of Section 3) is correct but it is again irrelevant for the idealized case and it is already published [5].

## 4. Some practical considerations about the *KLJN* cipher

The generic approach to address these and other kind of practical design/security problems has already been mentioned in [4] and a more detailed study [6] will soon be completed. Here, in advance, we show some of the generic issues of the practical design [6].

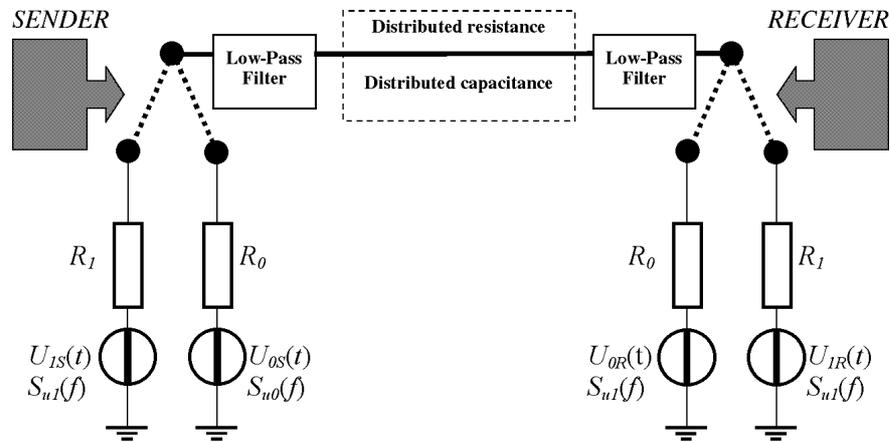

**Figure 1.** Simplest scheme of the *KLJN* cipher for practical design study [6].

Figure shows a simple approach to a practical design of the security of the *KLJN* cipher [6]. Eq. (1) above (same as Eq. (9) in [1]) is assumed as a governing rule, and low-pass filters at the line ends make it sure that this *no-wave condition* holds even if the noise has higher cut-off frequency, or if the switches are fast, or (most importantly) if the eavesdropper is trying to send high-frequency sampling signals into the line. The cable resistance issue is controlled by designing the system so that the combination of the relative voltage drop on the cable resistance and the clock duration do not provide sufficient statistics to guess the bits [4,6]. The cable capacitance is also an issue because voltage and current correlation measurements may cause information leak. The solution of the cable capacitance problem is similar to that of the cable resistance problem [6].

**Acknowledgments**

Discussions with a number of security experts and scientists/engineers challenging the security of the practical *KLJN* scheme are appreciated. Without the ability to provide a complete list, here are some of the most significant commenters: Derek Abbott, Steven M. Bellovin, Janos Bergou, Bob Biard, Terry Bollinger, Adrian Cho, David Deutsch, Julio Gea-Banacloche, Peter Heszler, Tamas Horvath, Greg Kochanski, Frank Moss, Rainer Plaga, Charlie Strauss, Olivier Saidi, Bruce Schneier, Matthew Skala, Charlie Strauss, Christian Vandenbroeck, David Wagner, Jonathan Wolfe.